\documentclass[preprint,prd,aps,superscriptaddress,showpacs,showkeys,nofootinbib,notitlepage ]{revtex4-1}
\usepackage{graphicx}
\usepackage{amsmath}
\usepackage{longtable}
\usepackage{color}

\newcommand{\ie}{i.e.,\,}
\newcommand{\be}{\begin{equation}}
\newcommand{\ee}{\end{equation}}
\newcommand{\bea}{\begin{eqnarray}}
\newcommand{\eea}{\end{eqnarray}}
\newcommand{\etal}{et al.}

\newcommand{\pht}{\tilde{\phi}_1}

\newcommand{\tp}{{\tilde{\phi}_1}}

\newcommand{\sit}{\sin\tilde{\phi}_1}

\newcommand{\sitsq}{\sin^2\tilde{\phi}_1}

\newcommand{\sitqu}{\sin^4\tilde{\phi}_1}

\newcommand{\cst}{\csc\tilde{\phi}_1}

\newcommand{\cstsq}{\csc^2\tilde{\phi}_1}
\newcommand{\cstcu}{\csc^3\tilde{\phi}_1}

\newcommand{\cstqu}{\csc^4\tilde{\phi}_1}

\newcommand{\cto}{\cos\tilde{\phi}_1}

\newcommand{\ctosq}{\cos^2\tilde{\phi}_1}
\newcommand{\ctocu}{\cos^3\tilde{\phi}_1}

\newcommand{\ctt}{\cot\tilde{\phi}_1}
\newcommand{\cttsq}{\cot^2\tilde{\phi}_1}
\newcommand{\Lrd}{\Lambda r_d^2}
\newcommand{\rsd}{\frac{r_s}{r_d}}

\newcommand{\logc}{\log\left[\cot\frac{\pht}{2}\right]}

\newcommand{\chir}{\frac{\chi_b}{\chi_d}}
\newcommand{\chirsq}{\left(\frac{\chi_b}{\chi_d}\right)^2}

\newcommand{\Ldfr}{\frac{\Lambda r_d^2}{3}}
\newcommand{\bd}{\beta_d}
\newcommand{\sd}{\frac{r_s}{r_d}}
\newcommand{\sdsq}{\left(\frac{r_s}{r_d}\right)^2}
\newcommand{\bdelt}{\mbox{\boldmath$\delta$}}
\newcommand{\tE}{\left(\frac{\theta_E D_d}{r_d}\right)}

\begin{document}

\bibliographystyle{apsrev}

\title{Image Properties of Embedded Lenses}

\author{R. Kantowski}
\email{kantowski@ou.edu}
\affiliation{Homer L.~Dodge Department~of  Physics and Astronomy, University of
Oklahoma, 440 West Brooks,  Norman, OK 73019, USA}

\author{B. Chen}
\email{bchen@ou.edu}
\affiliation{Homer L.~Dodge Department~of  Physics and Astronomy, University of
Oklahoma, 440 West Brooks,  Norman, OK 73019, USA}

\author{X. Dai}
\email{xdai@ou.edu}
\affiliation{Homer L.~Dodge Department~of  Physics and Astronomy, University of
Oklahoma, 440 West Brooks,   Norman, OK 73019, USA}

\date{\today}

\begin{abstract}
We give analytic expressions for image properties of objects seen around point mass lenses embedded in a flat $\Lambda$CDM universe.
An embedded lens in an otherwise homogeneous universe offers a more realistic representation of the lens's gravity field and its associated deflection properties than does the conventional linear superposition theory. Embedding reduces the range of the gravitational force acting on passing light beams thus altering all quantities such as  deflection angles,  amplifications,  shears and Einstein ring sizes.
Embedding also exhibits the explicit effect of the cosmological constant on these same lensing quantities.
In this paper we present these new results and demonstrate how they can be used.
The effects of embedding on image properties, although small i.e., usually less than a fraction of a percent,  have a more pronounced effect on image distortions in weak lensing where the effects can be larger than 10\%. Embedding also introduces a negative surface mass density for both weak and strong lensing, a quantity altogether absent in conventional Schwarzschild lensing.
In strong lensing  we find only one additional quantity, the potential part of the time delay, which differs from conventional lensing by as much as 4\%, in agreement with our previous numerical estimates.

\end{abstract}



\maketitle

\section{Introduction}

Recently we have investigated the quantitative effect of embedding on gravitational lensing observations by resorting to a mixture of analytic work with a few numerical applications  \cite{Kantowski10,Chen10,Chen11}.
The analytic results for quantities like the bending angle $\alpha$ produced by a point mass were given as functions of two impact variables $r_0$ and $\pht$ (see Fig.1).
These two parameters are not independent if the source and deflector redshifts are fixed.
Because of the non-linearity of the expressions we were only able to give an iterative procedure that allowed us to numerically evaluate  the conventional minimum impact Schwarzschild coordinate $r_0$ as a function of $\pht$ \cite{Chen11}.
We have since been able to analytically carry out this iterative procedure (see Eq.\,(\ref{r0}) in the appendix) and hence obtain all lensing properties such as position, shear, etc.,  as functions of the single impact angle $\pht$.
The solution of the embedded lens equation and comparison with classical lensing theory is therefore greatly simplified.
Because the dependence of observable quantities on this angle is highly nonlinear, we are not able to eliminate $\pht$ in favor of  $r_0$.
Derivations of our current results follow the steps given in \cite{Kantowski10,Chen10,Chen11} which we will not repeat but we will instead simply present the new results and use them on two examples.

The  point mass lens is the simplest lens to use to demonstrate the effects of embedding; however, all lenses will require corrections. An embedded point mass lens is constructed by condensing a comoving sphere of pressureless dust of a standard homogeneous cosmology to a singular point mass m at the sphere's center, a construction first made by Einstein himself \cite{Einstein45,Schucking54,Kantowski69,Kantowski95}. When the cosmology contains a cosmological constant $\Lambda$ the gravity field inside the evacuated sphere is described by the Kottler metric \cite{Kottler18,Dyer74} rather than the Schwarzschild metric. In this paper we restrict ourselves to a flat background cosmology whose Friedman-Lema\^itre-Robertson-Walker (FLRW) metric is

\be
ds^2=-c^2dT^2+R(T)^2\left[{d\chi^2}+\chi^2(d\theta^2+\sin^2\theta d\phi^2)\right].
\label{FLRW}
\ee
The embedded condensation is described by the Kottler or Schwarzschild-de Sitter metric
 \be
 \label{Kottler}
ds^2=-\gamma(r)^{-2}c^2dt^2+
\gamma(r)^2dr^2+
r^2(d\theta^2+\sin^2\theta\, d\phi^2),
\ee
where $\gamma^{-1}(r)\equiv\sqrt{1-\beta^2(r)}$ and $\beta^2(r)\equiv r_s/r+\Lambda r^2/3$.
The constants  $r_s$ and $\Lambda$  are the Schwarzschild radius  ($2G{\rm m}/c^2$) of the condensed mass and the cosmological constant respectively.
By matching the first fundamental forms  at the Kottler-FLRW boundary, angles $(\theta,\phi)$ of equations (\ref{FLRW}) and (\ref{Kottler}) are identified and the expanding Kottler radius $r_b$ of the void is related to its comoving FLRW boundary $\chi_b$ by
\be\label{rb-T}
r_b=R(T)\chi_b.
\ee
By matching the second fundamental forms the comoving FLRW radius $\chi_b$ is related to the Schwarzschild radius $r_s$ of the Kottler condensation by
\be\label{second-form}
r_s=\Omega_{\rm m}\frac{H_0^2}{c^2}(R_0\chi_b)^3.
\ee
Here  $H_0$ is the familiar Hubble constant and the cosmological constant $\Lambda$ is constrained to be the same inside and outside of the Kottler hole.

\begin{figure*}
\includegraphics[width=0.8\textwidth,height=0.26\textheight]{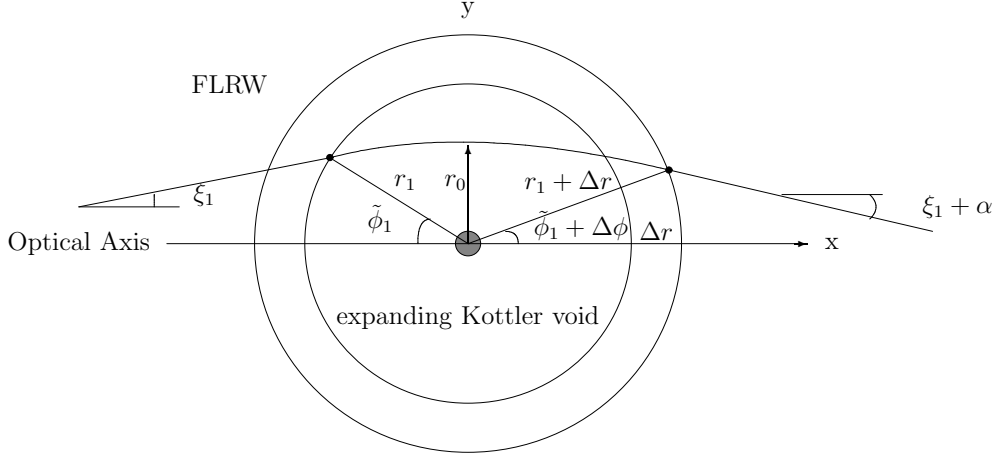}
\caption{A photon travels left to right entering a Kottler hole at $r=r_1,\, \phi=\pi-\tilde{\phi}_1$
and returns  to the FLRW dust at $r=r_1+\Delta r,\, \phi=\tilde{\phi}_1 +\Delta\phi$.
The photon's orbit has been chosen symmetric in Kottler about the point of closest approach $r=r_0$, $\phi=\pi/2$. Due to the cosmological expansion, $\Delta r>0$.
The slope of the photon's co-moving trajectory in the x-y plane is $\xi_1$
when incoming and $\xi_1+\alpha$ after exiting.
The resulting deflection angle as seen by a comoving observer in the FLRW background is $\alpha$, which is negative by convention.
Expressions for $r_1,$ $\Delta r,$ $\xi_1,$ $\Delta\phi,$ and $\alpha$ as functions of the two impact parameters, $r_0$ and $\pht$, can be found in \cite{Kantowski10,Chen10,Chen11}.}
\label{fig:fig1}
\end{figure*}

\begin{figure*}
\includegraphics[width=1.0\textwidth,height=0.17\textheight]{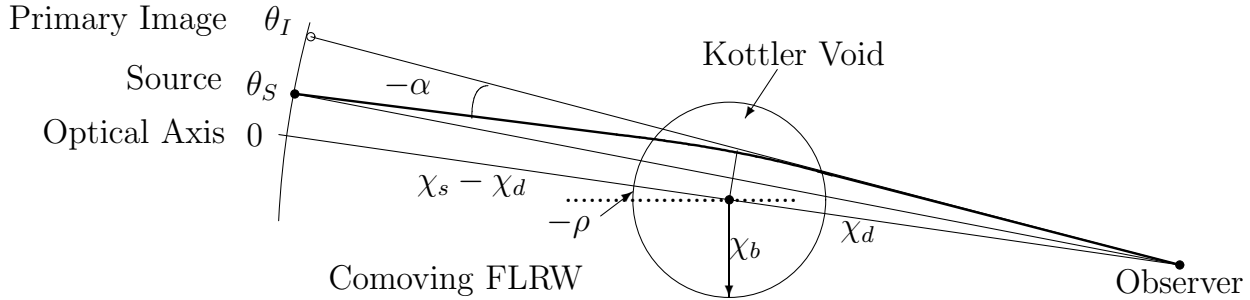}
\caption{A photon travels from a source at comoving distance $\chi_s$ from the observer and then enters a Kottler hole of comoving radius $\chi_b$ centered at comoving distance $\chi_d$ from the observer. The photon is deflected by an angle $\alpha <0$
and returns  to the FLRW dust on its way to the observer. Because this is a comoving picture the orbit inside the void is only representative and because the true orbit inside the void is symmetric about $\phi=\pi/2$ (see Fig. 1) the optical axis is rotated clockwise by the angle $-\rho$ (see Eq.\,(14) of \cite{Chen11}).}
\label{fig:fig2}
\end{figure*}

In the following sections we will give image locations and image properties of small sources seen through  Kottler voids in an otherwise flat FLRW universe (an embedded lens). We assume that the source and deflector are located at fixed FLRW comoving distances  $\chi_s$ and $\chi_d$  from the observer which correspond to angular diameter distances $D_s$ and $D_d$, and redshifts $1+z_s=R_0/R_s$ and $1+z_d=R_0/R_d$, see Fig.\,2. These quantities are computed just as if the void didn't exist. Any quantity with a subscript `$d$' means that it is evaluated at redshift $z_d$ when the radius of the universe was $R_d=R(T_d)=R_0/(1+z_d)$.  We give lensing properties such as the bending angle $\alpha$ of Eq.\,(\ref{alpha}) that are a series of smaller and smaller terms, sufficient to see both the shielding effect of embedding and the effect of the expansion rate $\beta_d=v_d/c$ of the void's Kottler radius $r_d=R_d\chi_b$ that existed at FLRW time $T_d$. The expansion rate $v_d$ is the speed of the expanding void boundary as measured by a stationary Kottler observer at $r_d$. It is given by evaluating $\beta(r)$ defined below Eq.\,(\ref{Kottler}) at $r=r_d$
\be
\beta_d=\sqrt{\sd+\Ldfr}.
\label{beta_d}
\ee
When expanding quantities such as $\alpha$ in a series we have taken parameters $\beta_d$ and $\chi_b/\chi_d=r_d/D_d$ (the angular radius of the Kottler hole,  see Fig.~\ref{fig:fig2}) to be first order and $r_s/r_d$ and $\Lambda r_d^2/3$ to be second order. In our results, e.g., Eq.\,(\ref{alpha}), we have used a parameter $\bdelt$ to keep track of each order. In Table 1 we give values for these and other parameters for two lens masses, ${\rm m}=10^{12}M_\odot$ (a large galaxy) and ${\rm m}=10^{15}M_\odot$ (a rich cluster) both at redshift $z_d=0.5$ in a flat $\Omega_{\rm m}=0.3$, $\Omega_\Lambda=0.7$ universe with $H_0=70\, \rm km\, s^{-1}\,Mpc^{-1}$. We refer to these as the galaxy lens and the cluster lens throughout the paper.

\begin{table}
\caption{\label{tab:parameters}Embedded lens parameters for two deflectors (a galaxy and a cluster) at $z_d=0.5$ when viewing a source at $z_s=1.0$ in a $\Omega_{\rm m}=0.3$, $\Omega_\Lambda=0.7$ universe with $H_0=70\, \rm km\, s^{-1}\,Mpc^{-1}$. Choosing $R_0=1$, gives  $\chi_d=1.89\times 10^{3}$ Mpc and $\chi_s=3.31\times 10^{3}$ Mpc.}
\begin{ruledtabular}
\begin{tabular}{cccccccc}
Lens & m & $\beta_d$ & $r_d/D_d=\chi_b/\chi_d$ & $r_s/r_d$ & $\Lambda r_d^2/3$ & $\theta_E$(rad) & $\tilde{\phi}_E$(rad) \\
galaxy& $10^{12}M_\odot$ & $3.67\times 10^{-4}$&$9.54\times 10^{-4}$&$7.98\times 10^{-8}$&$5.51\times 10^{-8}$&$8.07\times 10^{-6}$&$8.45\times 10^{-3}$\\
cluster& $10^{15}M_\odot$& $3.67\times 10^{-3}$&$9.54\times 10^{-3}$&$7.98\times 10^{-6}$&$5.51\times 10^{-6}$&$2.55\times 10^{-4}$&$2.65\times 10^{-2}$\\
\end{tabular}
\end{ruledtabular}
\end{table}

Shielding  typically causes the most significant embedding effect on images (i.e., the lowest order effect) and analytically appears as combinations of trig functions of the impact angle $\pht$ in quantities like the bending angle $\alpha$ (see the first $\ctocu $ term in Eq.\,(\ref{alpha}) below).   This decrease in $\alpha$ is caused by the shortened period a passing photon is influenced by the mass condensation (recall that in conventional lensing the deflecting force has $``\infty$'' range). Corrections caused by the presence of $\Lambda$ first appear in the void's expansion rate $\beta_d$ and are typically smaller than shielding corrections (i.e., are higher order). It was the search for $\Lambda$'s effect on light deflections \cite{Rindler07,Sereno08,Ishak08a,Ishak08b,Sereno09,Ishak10a,Ishak10b} that prompted investigations of embedded lensing \cite{Schucker09a,Schucker09b,Kantowski10,Boudjemaa}.

\section{The Embedded Lens Equation}

In our results we have introduced an order parameter  \mbox{\boldmath$\delta$} whose value is equal to 1 but whose purpose is to keep track of terms of similar orders as defined in the previous section. For weak lensing  impact angles $\pht$,  the higher the power of  \mbox{\boldmath$\delta$} the smaller the respective terms. For strong lensing, when  $\pht$ is sufficiently small, not all terms of a given order are of the same magnitude.
By using steps developed in \cite{Kantowski10,Chen10,Chen11} we find that to order $\mbox{\boldmath$\delta$}^4$ the source and image positions $\theta_S$ and  $\theta_I$, as functions of the single impact parameter $\pht$, can be written as
\bea
\theta_S&=&\theta_I+\frac{D_{ds}}{D_s}\,\alpha\Biggl\{1+\bdelt^2\frac{\chi_b}{2(\chi_s-\chi_d)}\times\cr
&&\left[\chir\sin^2\pht+\frac{2}{3}\bd\left(4-\sin^2\pht+3\sec\pht\log\left[\tan\frac{\pht}{2}\right]\right)\tan^2\pht \right]+{\cal O} \bigl(\bdelt^3\bigr)\Biggr\},
\label{theta_S}
\eea
(see Fig. 2), where the bending angle is
\bea
\alpha &=&-2\,\bdelt^2\sd\cst\Biggl\{\ctocu+\bdelt\Biggl[\bd\ctosq(1+2\sitsq)\Biggr]
+\bdelt^2\biggl[-\bd\chir\frac{1}{2}\ctocu\sitsq\cr
&+&\Ldfr\cto\sitsq(-1+4\sitsq)+\sd\Biggl(\frac{15}{16}\left(\frac{\pi}{2}-\pht\right)\cst-\ctt\cst\cr
&-&\frac{3}{2}\log\left[\cot\frac{\pht}{2}\right]\sitsq
+\cto\left(\frac{3}{16}+\frac{9}{8}\sitsq+\frac{13}{4}\sitqu \right)\Biggr)\Biggr]+{\cal O} \bigl(\bdelt^3\bigr)\Biggr\},
\label{alpha}
\eea
and
\bea\label{theta_I}
\theta_I&=& \bdelt\chir\sit\Biggl\{1- \bdelt\bd\cto+
\bdelt^2\Biggl[ \frac{1}{6}\chirsq\sitsq+\rsd\left(\cttsq -\frac{1}{2}\sitsq\right) +\Ldfr\left(1-\frac{3}{2}\sitsq\right)\Biggr]\cr
&+&\bdelt^3\Biggl[\chir\left(\frac{1}{4}\rsd-\frac{1}{2}\Ldfr\right)\cto\sitsq
-\bd\Biggl(\frac{1}{2}\chirsq\cto\sitsq + \Ldfr\cto\left(1-3\sitsq\right)\cr
&-&\sd\left(\cto\left(\frac{41}{12}-\frac{11}{12}\sitsq\right)+2\log\left[\tan\frac{\pht}{2}\right]\right)\Biggr)\Biggr]+{\cal O} \bigl(\bdelt^4\bigr)\Biggr\}.
\eea
We identify the pair of equations (\ref{theta_S}) and (\ref{theta_I}) above as the
embedded lens equation in parametric form. They can be used to obtain image positions and properties just as in conventional lensing theory. The conventional non-embedded lens equation is recovered by keeping only the lowest order terms in each expression and assuming $\cto \rightarrow 1$ and $r_d\sit\rightarrow r_0$.

\begin{figure*}
\includegraphics[width=0.6\textwidth,height=0.4\textheight]{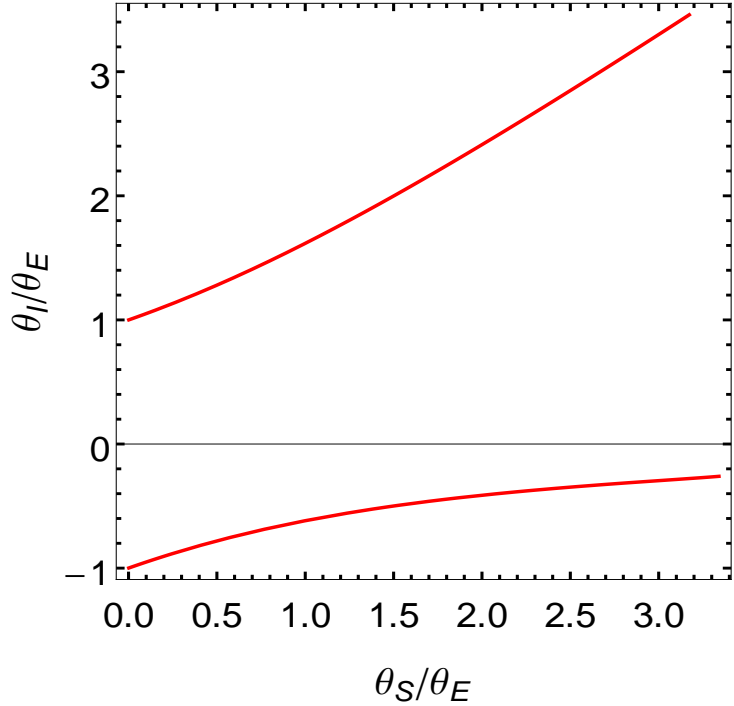}
\caption{ Primary and secondary image positions as functions of source position $\theta_S$ for the cluster lens of Table 1. Beyond $\theta_S\sim 3.4 \theta_E$ the smallness of the secondary image's impact begins to violate the orbit approximation condition  $\sin\pht \gg r_s/r_0$.}
\label{fig:fig3}
\end{figure*}

Because of the dependence of each $\bdelt$ order on the impact angle $\pht$, higher order terms that contain trig functions like $\cst$ or $\ctt$ can become comparable in magnitude with the next lower order terms for sufficiently small values of $\pht$.
This happens in strong lensing.
 For example the embedded Einstein ring size $\theta_E^\prime$  is found by first finding the value of
 $\pht=\tilde{\phi}_E$ that makes $\theta_S$ of Eq.(\ref{theta_S}) vanish (see $\tilde{\phi}_E$ values in Table 1) and then evaluating $\theta_E^\prime=\theta_I(\tilde{\phi}_E)$ using Eq.\,(\ref{theta_I}). For $\theta_S$ to vanish, $\bdelt$ and $\bdelt^2$ terms must cancel. The result is
 \be
\left(\theta^\prime_E\right)^2=\theta_E^2 \cos\tilde{\phi}_E^3\left(1+3 \bdelt\beta_d\tan\tilde{\phi}_E\sin\tilde{\phi}_E+{\cal O} (\bdelt^2)\right),
\label{theta_E^prime}
\ee
where $\theta_E$ is the conventional Einstein ring radius defined by
\be
\theta_E^2=\frac{2\,r_s D_{ds}}{D_d\,D_s}.
\ee
 As we have found with most strongly lensed image properties, this value differs only slightly from the conventional value. For the Einstein ring radius the embedded value differs somewhat more than 0.05\% for the cluster lens and 0.005\% for the galaxy lens.
In Figure 3 we have used the embedded lens equation to locate primary and secondary images for the cluster lens.
Primary and secondary images positions are given by Eq.\,(\ref{theta_I}) and correspond respectively to impact angles $\tilde{\phi}_{+}$ and $\tilde{\phi}_-$ (the Einstein impact angle $\phi_E$ separates the two image domains, \ie  $\tilde{\phi}_-<\phi_E<\tilde{\phi}_+$).
For a given source position $\Theta_S$, primary and secondary image impact angels $\tilde{\phi}_\pm$ are found by solving  $\theta_S(\tilde{\phi}_\pm)=\pm \Theta_S$ (i.e., by inverting Eq.\,(\ref{theta_S})).
The two images are then located at  $\pm\theta_I(\tilde{\phi}_\pm)$ (i.e., by using Eq.\,(\ref{theta_I})).
These two values of $\pht$ can then be used to determine primary and secondary image properties.

\section{Image Properties of The Embedded Lens}

To evaluate standard image properties the reader only has to compute the azimuthal and radial eigenvalues $(a_\phi,a_r)$ of the image matrix $\partial\boldsymbol\theta_S/\partial\boldsymbol\theta_I$ using equations (\ref{theta_S}) and (\ref{theta_I}).
We give them in Equations (\ref{aphi}) and (\ref{ar}) of the appendix. The primary and secondary values for   $a_\phi$ and $a_r$ can then be used to obtain image amplification, effective surface density, shear,  and eccentricity, respectively $\mu, \kappa, \gamma_s$, and $\epsilon$  (see \cite{Bourassa75,Ehlers}) by evaluating
\be
\mu^{-1}=a_\phi a_r,
\label{mu}
\ee
\be
\kappa=1-\frac{1}{2}(a_\phi+a_r),
\label{kap}
\ee
\be
\gamma_s=\frac{1}{2}(a_r-a_\phi),
\label{gam}
\ee
\be
\epsilon=\sqrt{1-\left(\frac{a_\phi}{a_r}\right)^2}.
\label{eps}
\ee
The above expressions give image properties for all values of impact angle $\pht$ such that the photon's orbit approximation is valid ($\sitsq\gg r_s/r_d$), but because of the lengths of the resulting  expressions, we find it appropriate to make two approximations in the next section, one for weak lensing and one for strong. The effective surface mass density $\kappa$ for the embedded lens is the one property that does not vanish as it does for the conventional Schwarzschild lens and is plotted in Figure\,4 for both weak and strong lensing of the primary cluster image. By a conventional Schwarzschild lens we mean conventional linear lensing theory applied to a point mass superimposed on a FLRW background. Even for strong lensing by the cluster the magnitude of $\kappa$ is only $\sim $ 0.1\% of the critical value. For the galaxy lens a $\kappa$ plot is similar to the cluster plot but is approximately a factor of 10 smaller.

\begin{figure*}
\includegraphics[width=0.6\textwidth,height=0.35\textheight]{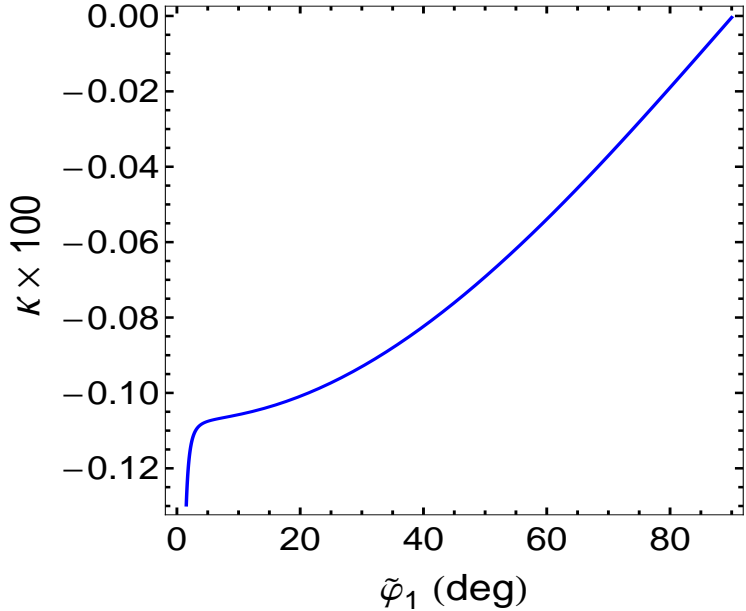}
\caption{The effective surface mass density $\kappa$ for the primary image of the embedded cluster lens of Table 1.}
\label{fig:fig4}
\end{figure*}

\section{Weak and Strong Approximations}

We found it necessary to keep terms to order $\bdelt^4$  in expressions such as $\alpha$ and $\theta_I$ to obtain sufficiently accurate results for most  strong lensing quantities.
Most weak observable quantities do not require such accuracy.
By dividing the domain for $\pht$ into  strong and weak parts we are able to give shorter expressions for the two eigenvalues  $(a_\phi,a_r)$  of equations (\ref{aphi}) and (\ref{ar}) and hence simpler expressions for $\mu,$ etc. For the strong domain we take  $0.4\phi_E<\pht<5\phi_E$ and for the weak $5\phi_E<\pht<\pi/2$.
The maximum value for $\theta_I$ is approximately the ratio $r_d/D_d$ which from Table 1 is $\sim$$117$ times the Einstein ring radius $\theta_E$ for the  cluster lens and $\sim$$37$ times for the galaxy.
Strong lensing consequently occurs for $\theta_I$ values up to $\sim$$5\theta_E$ and  weak lensing begins to occur when $\theta_I$ exceeds that value.
To obtain shortened expressions for weak lensing we need only keep terms of order $\bdelt^2$. This allows us to determine the lowest order effects of lens shielding and void expansion (the $\bd$ term) on image properties in the weak domain $5\phi_E<\pht<\pi/2$. We find that the approximate expressions are accurate to at least 0.1\%  down to $ \pht=5\,\phi_E$ for the cluster and to at least 0.03\% for the galaxy.
For weak lensing Eqs.~(\ref{aphi}) and (\ref{ar}) simplify to
\bea
a_\phi^{weak}&=&1-\bdelt\tE^2\cstsq\ctocu\Biggl\{1+\bdelt\bd\sec\pht(2+\sitsq)+{\cal O} \bigl(\bdelt^2\bigr)\Biggr\},\\
a_r^{weak}&=&1+\bdelt\tE^2\cstsq\Biggl\{\cto(1+2\sitsq)+\bdelt\bd(2-\sitsq+2\sitqu)
+{\cal O} \bigl(\bdelt^2\bigr)\Biggr\}.\nonumber
\label{arw}
\eea
From these the following image properties result:
\bea
(\mu^{weak})^{-1}&=&1+3\,\bdelt \tE^2\Biggl\{\cto+\, \bdelt \bd\sitsq+{\cal O} \bigl(\bdelt^2\bigr) \Biggr\}\cr
&&-\bdelt^2 \tE^4\cot^4\pht\Biggl\{(1+2\sitsq)
+{\cal O} \bigl(\bdelt\bigr)\Biggr\},\label{muw}\\
\kappa^{weak} &=&-\frac{3}{2}\, \bdelt \tE^2\Biggl\{\cto+\, \bdelt\, \bd\sitsq+{\cal O} \bigl(\bdelt^2\bigr) \Biggr\},\\
\gamma^{weak}&=&\bdelt \tE^2\cstsq\Biggl\{\cto\left(1+\frac{1}{2}\sitsq\right)+\,\bdelt  \bd\left(2-\sitsq+\frac{1}{2}\sitqu\right)+{\cal O} \bigl(\bdelt^2\bigr) \Biggr\},\hspace{19pt}\\
\epsilon^{weak}&=&
\Biggl\{\sqrt{\bdelt}\tE\cst\sqrt{2\cto(2+\sitsq)+\, \bdelt\, \bd(4-2\sitsq+\sitqu)+{\cal O} \bigl(\bdelt^2\bigr)}\Biggr\}\cr
&\times& 1/\Biggl\{1+\bdelt\tE^2\cstsq\Bigl[\cto(1+2\sitsq)+{\cal O} \bigl(\bdelt\bigr)\Bigr]\Biggr\}.
\label{epsilonw}
\eea

In Figure 5 we have compared image shear and ellipticity of the embedded lens images with  conventional (non-embedded) Schwarzschild values.
The reader can see that beyond $\pht\sim 45^\circ$ the embedded lens differs from Schwarzschild by over 10\%.
This is caused primarily by the shielding of the embedded mass and increases as the  transiting light ray's minimum impact $r_0$ approaches the void boundary.
The embedded amplification $\mu$ differs from conventional Schwarzschild by less than 0.2\% for the cluster and 0.02\% for the galaxy for the weak lensing domain and only increases to 2.5\% for the secondary cluster image in the strong lensing limit $\pht\rightarrow 0.4\phi_E$ where $\mu\rightarrow -0.03$.
The galaxy lens' numbers are significantly less and neither are plotted.

\begin{figure*}
\begin{center}$
\begin{array}{cc}
\includegraphics[width=0.52\textwidth,height=0.3\textheight]{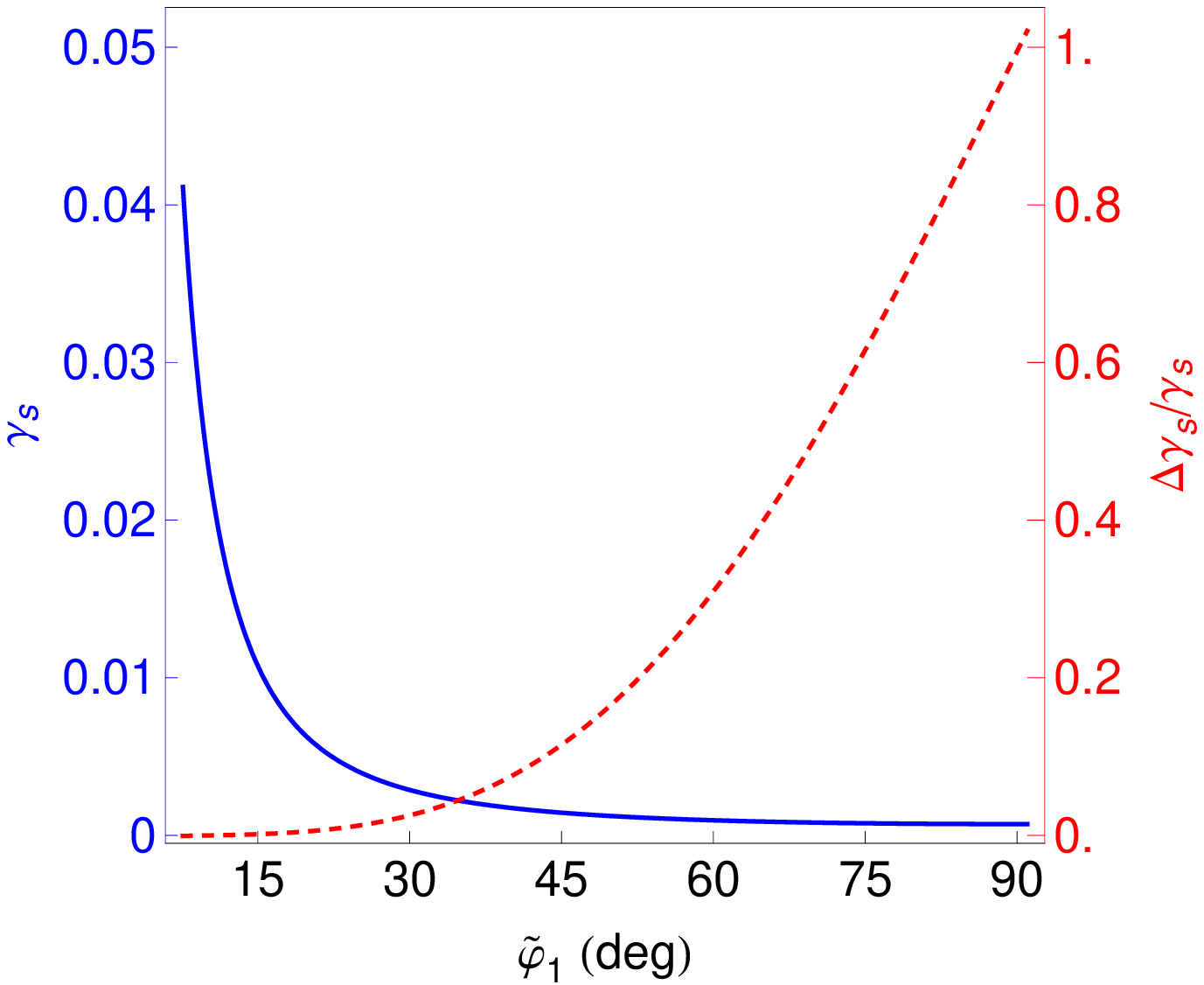}
\hspace{10pt}
\includegraphics[width=0.5\textwidth,height=0.3\textheight]{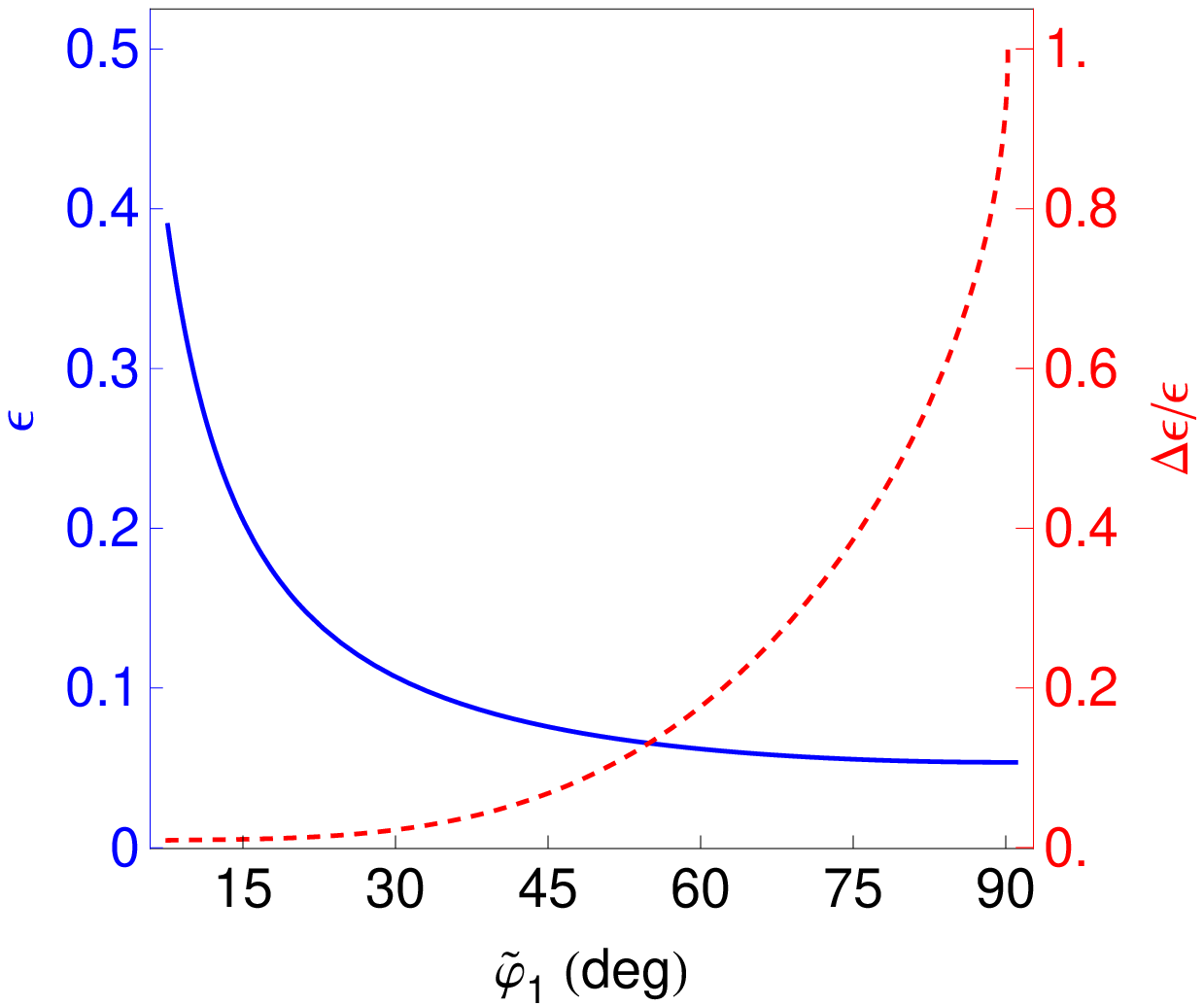}
\end{array}$
\end{center}
\caption{ Corrections to image shear (left panel) and ellipticity (right panel) caused by embedding of the cluster lens of Table 1.
The solid blue curves in the left and right panels are respectively the $\gamma_s$ and $\epsilon$ for the conventional (non-embedded) Schwarzschild lens.
The fractional difference in the image shear, $\Delta\gamma_s/\gamma_s$, and ellipticity, $\Delta\epsilon/\epsilon,$ caused by embedding, are the dashed red curves in the left and right panels, plotted as a functions of $\pht$.
Differences are computed at the same primary image positions $\theta_I(\pht).$
 }
\label{fig:fig5}
\end{figure*}

Using the order parameter $\bdelt$ to track terms of equal importance is problematic  for strong lensing.
As discussed above for strongly lensed images,  small values of $\pht$ in trig functions like $\cstsq$ increase the numerical magnitudes of some of the terms in Eqs.~(\ref{aphi}) and (\ref{ar}).
In the following strong lensing approximation we have kept terms based on their numerical size at the Einstein ring value $\pht=\phi_E$, and ordered them using another parameter $\Delta$ whose value is also 1.
For the cluster lens the $\Delta^1$ terms are of numerical order 0.1,  $\Delta^2$ terms are of numerical order 0.01 and so on. For the galaxy lens all terms are $\sim 1/10$ those of the cluster.
The principal eigenvalues $a_\phi$ and $a_r$ of Eqs.~(\ref{aphi}) and (\ref{ar}) are approximated by
\bea
a_\phi^{strong}&=&\Delta^2\Biggl[1-\tE^2\cstsq\Biggl\{\ctocu-2\,\sd\cstsq\cto+2\,\Delta\bd\ctosq+{\cal O}\bigl(\Delta^2\bigr)\Biggr\}\Biggr],\\
a_r^{strong}&=&1+\tE^2\Biggl\{\cstsq\cto-2\,\Delta^2\sd  \cstqu\cto+2\,\Delta^3(\cto+\bd \cstsq)+{\cal O}\bigl(\Delta^4\bigr)\Biggr\}.\nonumber
\label{ars}
\eea
These approximate expressions are accurate to at least 0.2\% for the strong domain $0.4\,\phi_E<\pht<5\,\phi_E$ for the cluster lens and accurate to 0.01\% for the galaxy.
Strong lensing image properties given in equations (\ref{mu})-(\ref{eps})  differ from conventional Schwarzschild values by only a fraction of a percent and are not separately approximated.
The effective surface mass density $\kappa$ of Eq.~(\ref{kap}), which no longer vanishes as it does for the non-embedded Schwarzschild lens, can be approximated to an accuracy of more than 0.01\% for the strong domain as
\be
\kappa^{strong}=-\frac{3}{2}\tE^2\Biggl(\cto+\frac{5\pi}{32}\sd\cstcu\Biggr).
\ee

An additional strong lensing property of importance is the time delay.
It contains a geometric part and a potential part, \ie $\Delta T=\Delta T|_g+\Delta T|_p$, see \cite{Ehlers,Cooke75,Schucker10a}.
The arrival time differences for the two images caused by the difference in geometrical path lengths for our embedded Swiss cheese (SC) lens  is $\Delta T_{\rm SC}|_g$ and when computed to maximum accuracy as described in \cite{Chen10}, proves to be almost indistinguishable from the conventional Schwarzschild value $\Delta T_{\rm Sch}|_g$
\be
\frac{c\Delta T_{\rm SC}|_g}{r_s}\approx\frac{c\Delta T_{\rm Sch}|_g}{r_s}=(1+z_d)\left(\frac{\theta_S}{\theta_E}\right)\sqrt{ \left(\frac{\theta_S}{\theta_E}\right)^2+4}.
\ee
The potential part of the embedded lens delay, $\Delta T_{\rm SC}|_p$, as defined in \cite{Chen10}, is given by taking the difference in the following for the primary and secondary images
\be
\frac{c\Delta T_p}{r_s}=2\,(1+z_d)\Biggl\{ \log\left[\cot\frac{\tp}{2}\right]-\cto\left(1+\frac{1}{3}\cos^2\pht\right)+\, \mbox{\boldmath$\delta$}\, \beta_d\cos^4\tilde{\phi}_1 +{\cal O} \bigl(\bdelt^2\bigr)\Biggl\}.
\label{T_p}
\ee

\begin{figure*}
\includegraphics[width=0.6\textwidth,height=0.36\textheight]{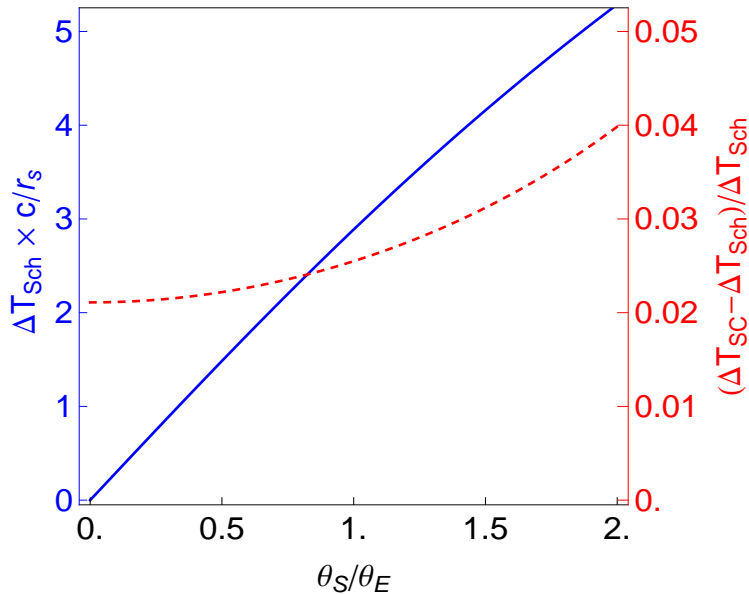}
\caption{A comparison of the potential parts of the time delay ($\Delta T_{\rm SC}|_p$) of the embedded cluster lens with the conventional theory.
The comparison is for sources at same positions even though the image positions for the two theories are different.
The solid blue line is the conventional potential part of the time delay and the dashed red line is the fractional difference of the embedded and the conventional theories. }
\label{fig:fig6}
\end{figure*}

In Figure 6 we have compared  $\Delta T_{\rm SC}|_p=\Delta T_p^{\rm secondary}-\Delta T_p^{\rm primary}$ (\ie the potential part of the time delay) of the cluster lens with the corresponding conventional (non-embedded) Schwarzschild value. The reader  can see that there is a 2--4\% difference in arrival times between the theories.

\section{Conclusions}

This paper is one of a series of investigations of the differences in image properties caused by including the gravitational lens's mass in the cosmic mean density.
We call such a lens an embedded lens.
In this paper we have eliminated one of the two impact parameters previously required to give embedded point mass lensing quantities such as the bending angle $\alpha$ and the lens equation itself.
The theory remains more complicated than the conventional lensing theory, but is now much easier to use.
The new analytical expressions for image properties agree  with the lowest order results given in \cite{Chen11}.
They can also be compared with the higher order results in \cite{Kantowski10,Chen10} that were given as functions of the two impact parameters $r_0$ and $\pht$.
To eliminate $r_0$ in our prior results for quantities such as $\alpha(\pht,r_0)$ in Eq.~(32) of \cite{Kantowski10} and obtain results such as Eq.\,(\ref{alpha}) given in this paper we had to analytically iterate Eq.~(17) of \cite{Chen11} to determine $r_1(\pht)$ and then use the orbit equation (11) of \cite{Kantowski10} to determine  $r_0(\pht)$.
The result is given in Eq.\,(\ref{r0}) of the appendix for completeness and to allow the reader to eliminate $r_0$ in other quantities of interest.

We have found that with the exception of the potential part of the time delay and the effective surface mass density $\kappa$, strong lensing quantities are only minimally altered by making the lens mass a contributor to the mean mass density of the universe. Even there the effect is less than 5\% on the time-delay for a huge cluster lens, see Fig.\,6. For weak lensing most effects are also small; however, shear and image ellipticity begin to differ significantly ($>10$\%, see Fig.\,5) for large impact angles $\pht>45^\circ$. The one quantity that doesn't vanish in embedded point mass lensing is $\kappa$. It turns out to be negative, presumably accounting for the missing FLRW mass density in the Kottler void.

All results given here depend on having a flat ($\Omega=1$)  background. Extending them to $\Omega\ne1$ is clearly possible. We expect that many results will differ trivially from what we have given here. The applicability of all results given here also depends on the lens being sufficiently condensed so as to be approximated by a point mass. The effects of embedding on extended lenses remains to be investigated \cite{Schucker10b}.

To correct for embedding we have used the Swiss cheese cosmologies which are commonly criticized for their unrealistic mass distributions, i.e., holes with masses at their centers that abruptly appear in otherwise uniform backgrounds.
The abrupt discontinuity that appears in the cheese is certainly an unrealistic representation of the true matter distribution; however, this is primarily an aesthetic complaint.
Fortunately for Swiss cheese, its purpose is not to represent the mass distribution but instead to account for the  effects of mass inhomogeneities on the local/global dynamics of the geometry and on the optics of transiting light rays.
In those two aspects Swiss cheese does quite well.
The real shortcoming of a simple Swiss cheese type embedded lens (a single condensation moving with the Hubble flow) is the absence of any shear at the site of the embedded lens.
For such a simple embedded lens, neighboring inhomogeneities can only be distributed so as to produce a homogenized gravity field at the lens site.
Consequently the accuracy of our predictions can be questioned. Stated simply, the shortcoming of our lens model, and with standard Swiss cheese itself,  is that  neighboring  and distant inhomogeneities produce an homogenized background at the point where the lens inhomogeneity is inserted.
We suspect this ``average''  lens is not representative because it does not account for effects of local shear.
We currently do not have a good estimate of how much de-homogenization  alters the shielding radius (which is the major source of embedding effects) because there are no simple Einstein solutions which accurately model local distortions.
Such distortions can easily be accommodated in conventional lensing theory, but how they would alter the embedding radius is completely unknown.
Exact Einstein solutions containing a local shear can be constructed by using hierarchical models built from Swiss cheese itself.
Such a construction will probably be necessary to dependably estimate how the spherical shielding radius $r_d$ is distorted and possibly extended by a local shear and hence how it modifies predictions made here.

\begin{acknowledgments}
NSF AST-0707704, and US DOE Grant DE-FG02-07ER41517 and Support for Program number HST-GO-12298.05-A was provided by NASA through a grant from the Space Telescope Science Institute, which is operated by the Association of Universities for Research in Astronomy, Incorporated, under NASA contract NAS5-26555.
\end{acknowledgments}

\appendix*
\section{}

The minimum Kottler radial coordinate $r_0$ as a function of impact angle $\pht$ (see Fig.\,1) is
\bea\label{r0}
r_0&=&r_d\sit\Biggl\{1-\bd\bdelt\cto\crcr
&+&\bdelt^2\Biggl[\bd\chir\frac{1}{2}\sitsq+\rsd\left(\cstsq-\frac{1}{2}\cst-\frac{1}{4}-\frac{5}{4}\sitsq\right)+\Ldfr(1-2\sitsq)\Biggr]\cr
&+&\bdelt^3\Biggl[-\chir\left(\frac{5}{4}\rsd+2\Ldfr\right)\cto\sitsq+\bd\Biggl(\Ldfr\cto\left(-1+\frac{9}{2}\sitsq\right)\cr
&&+\rsd\left(2\log\left[\tan\frac{\pht}{2}\right]+\cto\left(\frac{5}{3}+\frac{7}{3}\sitsq\right)\right)\Biggr)\Biggr]\cr
&+&\bdelt^4\Biggl[\left(\Ldfr\right)^2\left(1-10\sitsq+11\sitqu\right)+\sdsq\Biggl(\frac{15}{16}\left(\frac{\pi}{2}-\tp\right)\ctt\cstsq\cr
&&-\cto\log\left[\tan\frac{\pht}{2}\right]-\cstqu+\frac{17}{16}\cstsq-\frac{187}{48}-\frac{4}{3}\sitsq+\frac{59}{12}\sitqu\Biggr)\cr
&&+\bd\chir\left(\frac{1}{8}\chirsq\sitqu+\Ldfr\left(5\sitsq-7\sitqu\right)+\sd\left(1-\frac{3}{2}\sitqu\right)\right)\cr
&&-\sd\Ldfr\left(4\cto\log\left[\tan\frac{\pht}{2}\right]+\frac{137}{24}+\frac{103}{12}\sitsq-\frac{385}{24}\sitqu\right)\cr
&&+\frac{1}{4}\chirsq\left(\frac{1}{4}\sd+\Ldfr\right)\sitqu \Biggr]+{\cal O} \bigl(\bdelt^5\bigr)\Biggr\}.
\eea
All quantities such as $\xi_1$, $\Delta\phi,$ $\Delta r,$ and $\rho$ (see Fig.\,1 and Fig.\,2), previously given as functions of $\pht$ and $r_0$ \cite{Kantowski10,Chen10,Chen11} can be expressed as functions of the single impact parameter $\pht$ using Eq.\,(\ref{r0}).

The azimuthal and radial (with respect to the optical axis,  see Fig.\,2) eigenvalues $(a_\phi,a_r)$ of the lensing matrix $ \partial \mbox{\boldmath$\theta$}_s/\partial \mbox{\boldmath$\theta$}_I$  to order $\bdelt^3$ as functions of the impact angle $\pht$ and an additional lens-geometry parameter $(\theta_ED_d/r_d)^2$ (a term which is of order $\bdelt$) are
\bea
a_\phi&=&1-\bdelt\tE^2\Biggl\{\ctocu\cstsq\left(1+\bdelt\bd\sec\pht(\sitsq+2)\right)
+\bdelt^2\Biggl[\chirsq\frac{4\chi_d-\chi_s}{6(\chi_s-\chi_d)}\ctocu\cr
&&+\sd\Biggl(\cto\cstqu\left(-2+\frac{67}{16}\sitsq-\frac{3}{8}\sitqu+\frac{7}{4}\sin^6\pht \right)+\frac{15}{16}\left(\frac{\pi}{2}-\pht\right)\cstcu-\cr
&&\frac{3}{2}\log\left[\cot\frac{\pht}{2}\right]\Biggr) -\bd\frac{\chi_b}{\chi_s-\chi_d}\left(\frac{1}{2}\ctocu\frac{\chi_s-\chi_d}{\chi_d}-\log\left[\tan\frac{\pht}{2}\right]-\frac{1}{3}\cto\left(4-\sitsq\right) \right) \cr
&&+\frac{\Lrd}{3}\cto\cstsq\left(1+\frac{1}{2}\sitsq+\frac{3}{2}\sitqu\right) \Biggr]+{\cal O}\left(\bdelt^3\right)  \Biggr\},
\label{aphi}
\eea
and
\bea
a_r&=&1+\bdelt\tE^2\Biggl\{\cto\cstsq(1+2\sitsq)+\bdelt\bd\cstsq(2-\sitsq+2\sitqu)\cr
&&+\bdelt^2\Biggl[ \chir\frac{\chi_b}{(\chi_s-\chi_d)}\cto\left(3\sitsq-\frac{1}{2}\frac{\chi_s}{\chi_d}(1+2\sitsq)\right)+\frac{1}{2}\bd\frac{\chi_b}{(\chi_s-\chi_b)}\cto\times\cr
&&\Biggl(\frac{\chi_s}{\chi_d}(1-4\sitsq)+2\sec\pht\logc-\frac{1}{3}(17-20\sitsq)\Biggr)  +\frac{1}{32}\rsd\cstqu\times\cr
&&\Biggl(\cto(-64+164\sitsq-64\sitqu-192\sin^6\pht)+48\sitqu\logc\cr
&&+60\left(\frac{\pi}{2}-\pht\right)\sit\Biggr) +\frac{\Lrd}{6}\cto\cstsq(2+\sitsq-6\sitqu)
 \Biggr]+{\cal O}\left(\bdelt^3\right)  \Biggr\}.
\label{ar}
\eea

\label{lastpage}


\begin{thebibliography}{breitestes Label}

\bibitem[Kantowski \etal (2010)]{Kantowski10} R. Kantowski, B. Chen  \& X. Dai,  \apj, 718, 913 (2010).

\bibitem[Chen \etal (2010)]{Chen10} B. Chen, R. Kantowski  \& X.  Dai,  \prd, 82, 043005 (2010).

\bibitem[Chen \etal (2011)]{Chen11} B. Chen, R. Kantowski  \& X.  Dai,  \prd, 84, 083004 (2011).

\bibitem[Einstein \& Straus(1945)]{Einstein45} A. Einstein \& E. G. Straus, Rev. Mod. Phys., 17, 120 (1945).

\bibitem[Sch\"ucking(1954)]{Schucking54} E. Sch\"ucking,  Z. Phys., 137, 595 (1954).

\bibitem[Kantowski(1969)]{Kantowski69} R. Kantowski,   \apj, 155, 89 (1969).

\bibitem[Kantowski, Vaughan, \& Branch(1995)]{Kantowski95} R. Kantowski, T. Vaughan \& D. Branch,  \apj, 447, 35 (1995).

\bibitem[Kottler(1918)]{Kottler18} F. Kottler, Ann. Phys. (Leipzig), 361, 401 (1918).

\bibitem[Dyer \& Roeder(1974)]{Dyer74} C. C. Dyer \& R. C. Roeder,  \apj, 189, 167 (1974).

\bibitem[Rindler \& Ishak(2007)]{Rindler07} W. Rindler \& M. Ishak, \prd, 76, 043006 (2007).

\bibitem[Ishak et al.(2010)]{Ishak10a} M. Ishak, W. Rindler \& J. Dossett, Mon. Not. R. Astron. Soc., 403, 21521 (2010).

\bibitem[Ishak \& Rindler(2010)]{Ishak10b} M. Ishak \& W. Rindler,   Gen. Relativ. Gravit., 42, 2247 (2010).

\bibitem[Sereno (2009)]{Sereno09}  M. Sereno,  \prl, 102, 021301 (2009).

\bibitem[Sereno (2008)]{Sereno08} M. Sereno,  \prd, 77, 043004 (2008).

\bibitem[Ishak(2008)]{Ishak08a} M. Ishak,  \prd, 78, 103006 (2008).

\bibitem[Ishak et al.(2008)]{Ishak08b} M. Ishak, W. Rindler, J. Dossett, J. Moldenhauer \& C. Allison,  Mon. Not. R. Astron. Soc., 388, 1279 (2008).

\bibitem[Sch\"{u}cker(2009a)]{Schucker09a} T. Sch\"ucker, Gen. Relativ. Gravit., 41, 67 (2009).

\bibitem[Sch\"{u}cker(2009b)]{Schucker09b} T. Sch\"ucker,  Gen. Relativ. Gravit., 41, 1595 (2009).

\bibitem[Boudjemaa et al.(2011)]{Boudjemaa} K.-E. Boudjemaa, M. Guenouche  \& S. R. Zouzou,  Gen. Relativ. Gravit., 43, 1707 (2011).

\bibitem[Bourassa \& Kantowski (1975]{Bourassa75} R. R. Bourassa \& R. Kantowski, Ap. J. {\bf 195}, 13 (1975).

\bibitem[Schneider \etal (1992)]{Ehlers} P. Schneider, J Ehlers \& E. E.  Falco,  {\it Gravitational Lenses} (Springer-Verlag, Berlin, 1992).

\bibitem[Cooke \& Kantowski (1975)]{Cooke75} J. H. Cooke and  R. Kantowski,  Ap. J. {\bf 195}, L11 (1975).

\bibitem[Sch\"{u}cker(2010a)]{Schucker10a}  T. Sch\"ucker, arXiv:1006.3234 (2010).

\bibitem[Sch\"{u}cker(2010b)]{Schucker10b} T. Sch\"ucker,  Gen. Relativ. Gravit., 42, 1991 (2010).


\end{thebibliography}
\end{document}